\documentclass[RNAAS]{aastex631}

\graphicspath{{./}{figures/}}

\begin{document}

\title{Disentangling Joint Light Curves Using Photocenter Shifts: A Case Study Using NEOWISE Data}

\correspondingauthor{Bringfried Stecklum}
\email{stecklum@tls-tautenburg.de}

\author[0000-0001-6091-163X]{Bringfried Stecklum}
\affil{Thüringer Landessternwarte Tautenburg,
Sternwarte 5,
Tautenburg, 07778, Germany}

\begin{abstract}
The data collected by the Wide-field Infrared Survey Explorer (WISE, \citealt{2010AJ....140.1868W}) and its follow-up Near Earth Object (NEO) mission (NEOWISE, \citealt{2011ApJ...743..156M}) represent a treasure trove for variability studies. However, the angular resolution imposed by the primary mirror implies that close double stars are often unresolved. Then, variability of one or both stars leads to motion of the image centroid along the connecting line. 
Knowledge of the angular separation, derived from higher-resolution imaging which resolves both components, allows disentangling of the joint light curve into individual ones. This is illustrated by the case of \object{SPICY 1474} which featured an outburst several years ago. Removal of the contribution of the nearby companion, which led to a dilution of the burst strength, revealed that it was $\sim$0.5\,mag brighter than in the joint light curve. A comparison with light curves from unTimely suggests that utilizing the photocenter shift may lead to more reliable results. 
\end{abstract}

\section{Introduction} \label{sec:intro}

The cryogenic WISE space telescope, launched in 2009, featured a 40-cm mirror to perform all-sky surveying in the mid-infrared. It was hibernated in 2011 after coolant depletion and reactivated in 2013 for NEOWISE, primarily to detect near-Earth asteroids. Its Sun-synchronous orbit implied a sampling time of half a year. Ultimately, (NEO)WISE was decommissioned in mid-2024.
Though the time sampling is sparse, the 14-year baseline makes (NEO)WISE data invaluable for variability studies.
Numerous such investigations have been performed for young stellar objects (YSOs), e.g. 
\cite{2022ApJ...924L..23Z}, \cite{2024ApJ...962...38L}.
A particular objective is to identify and monitor episodic accretion bursts which represent short but intense episodes of protostellar growth, e.g. 
\cite{2020MNRAS.499.1805L}, \cite{2023ApJ...957....8W}. 

Along these lines, the recent work of \cite{2025arXiv250313971S} classified 235 sources as bursters. In order
to validate this classification, 
I retrieved and inspected their (NEO)WISE light curves.
During this task, one YSO, \object{SPICY 1474} \citep{2021ApJS..254...33K}, caught my attention due to a significant linear shift of the photocenter over time. Centroid shifts can occur in YSOs due to varying illumination of their reflection nebulae, e.g., \cite{2024A&A...682A..17S}, but this is absent in \object{SPICY 1474}. The existence of a nearby companion, \object{SPICY 1478}, causes both to appear as an unresolved binary star to (NEO)WISE. Its position is that of the photocenter, i.e. the flux-weighted mean position of both components,
which shifts toward the brightening star and away when dimming. According to astrometric terminology, the apparent binary star SPICY 1474/1478 thus represents
a ``variability-induced mover'' (VIM, \citealt{1996A&A...314..679W}).
Resolving the VIM allows for the light curve to be divided into its components using centroid shifts and known separation.

\section{Data retrieval and analysis}
(NEO)WISE photometry and positions for the presumed bursters have been retrieved from the NASA/IPAC Infrared Science Archive (IRSA) within 5\arcsec{} of the YSO positions, based on frames with the photometric quality flag A. In addition, ALLWISE and \emph{Spitzer} IRAC images for \object{SPICY 1474} were obtained via the IRSA viewer service.
The fluxes and coordinates 
from the NEOWISE-R Single Exposure (L1b) Source Table 
were averaged for each observing epoch, yielding mean values and corresponding errors. 
The separation between \object{SPICY 1474} and \object{SPICY 1484} as inferred from the GAIA positions amounts to 4\farcs67. For comparison, the full widths at half-max of the W1 (3.4\,\micron) and W2 (4.6\,\micron) images amount to 8\farcs5 for ALLWISE and 6\arcsec{} for unblurred coadds \citep{2014AJ....147..108L}.

The distance between the centroid and one component is inversely proportional to its brightness.
To derive the flux for one component, the measured flux was multiplied by the respective distance of the {\em other} component from the observed centroid position, divided by the separation. 
The implicit assumption made here is that the distance between the two components remains constant.

For comparison, light curves were also established using unTimely \citep{2023AJ....165...36M}.
This tool, which performs photometry on unblurred (NEO)WISE images \citep{2014AJ....147..108L} has been used to search for YSOs showing large-amplitude variations as well \citep{2024MNRAS.532.2683L}. The unTimely light curves were generated by forced photometry with a box size of 2\arcsec{} for the positions of the two objects. Currently, unTimely covers 16 (NEO)WISE epochs until 2021.

\section{Results and conclusions}

\begin{figure}[h]   
\centering
\includegraphics[width=\textwidth]{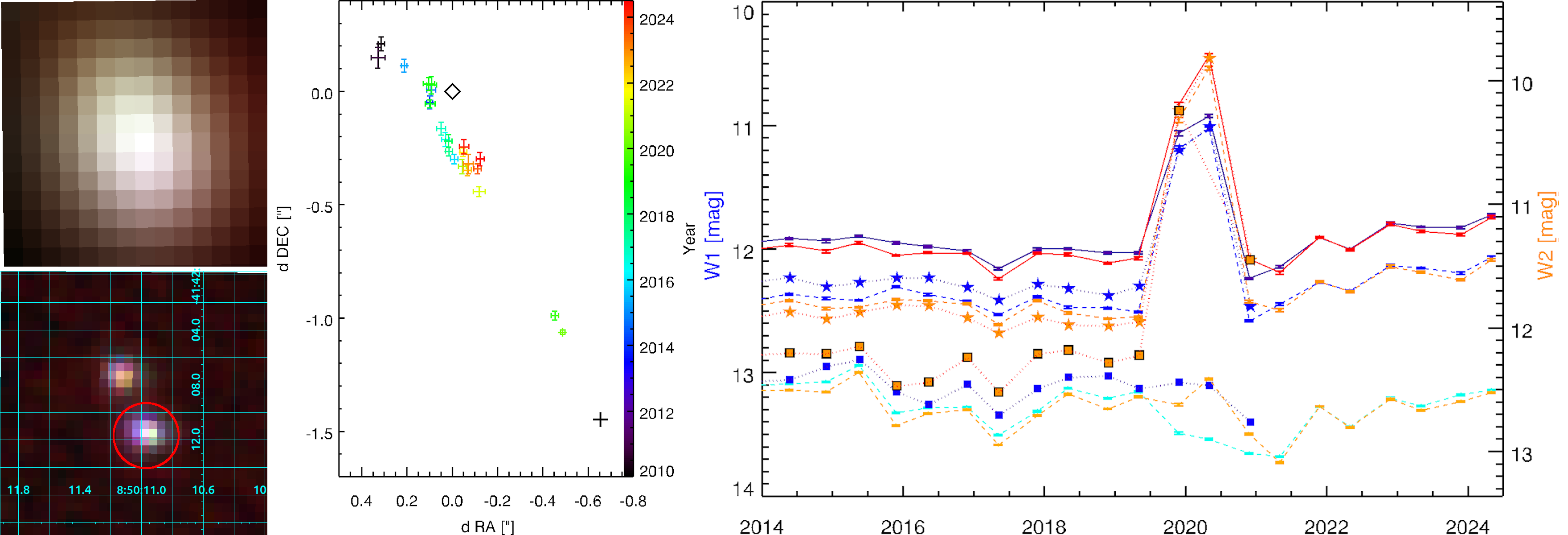}
\caption{{\bf Left:} RGB composites (field of view 20\arcsec{}$\times$20\arcsec{}) based on images from ALLWISE (top) and IRAC (bottom). \object{SPICY 1474} is marked by the red circle.
{\bf Center:} Centroid positions relative to that resulting from GAIA DR2 positions and IRAC I1 fluxes (diamond symbol). The cross marks the \object{SPICY 1474} position while that of \object{SPICY 1478} is outside the field. (NEO)WISE epochs are color-coded. {\bf Right:} NEOWISE light curves: blue - W1, red - W2;  dashed line - \object{SPICY 1474}, light dashed line - \object{SPICY 1478}, full line - joint. The unTimely values have filled symbols (asterisks - \object{SPICY 1474}, squares - \object{SPICY 1478}). The unTimely W2 light curve of \object{SPICY 1478} is marked by black-bordered symbols since it deviates from the disentangled one.
}
\label{fig:lcd}
\end{figure}

Figure \ref{fig:lcd} shows the main result of the study. The left side demonstrates that (NEO)WISE could not resolve the components, unlike {\em Spitzer} IRAC. The adjacent panel shows the centroid distribution over time relative to IRAC observations, aligning with the apparent double star's position angle. The photocenter's shift towards \object{SPICY 1474} is due to its brightening. The right panel shows the W1 and W2 NEOWISE light curves. The joint light curves for W1 and W2 established from the L1b table are at the top. The disentangled light curves for \object{SPICY 1474} reveal that the burst peak is slightly dimmer than the joint light curves, while the quiescent flux is nearly half a magnitude lower. The removal of the contribution from the companion led to a relative increase in the burst strength, which is crucial to assess its energetics. The disentangled curves for the fainter \object{SPICY 1478} are at the bottom.

Fig.\,\ref{fig:lcd} also shows that the UnTimely light curves of \object{SPICY 1474} agree quite well with the disentangled ones. However, for \object{SPICY 1478}, the W2 unTimely light curve deviates from the disentangled one during the burst of \object{SPICY 1474} and shows an elevated flux level before. This can be attributed to the slightly wider point spread function (PSF) in the W2 band, which incurs a greater light spillover at the separation of the two YSOs.

The burst sample of \cite{2025arXiv250313971S} includes 24 additional objects with an elongated centroid distribution, representing 10\% of the sample size, prompting investigation into whether unresolved companions are the cause. The burst of \object{SPICY 1474} was also detected by GAIA (Gaia19fkq, \citealt{2021A&A...648A..44M}) and ATLAS \citep{tonryATLASHighcadenceAllsky2018}. The ATLAS light curve shows an unusual double-peaked profile. A more comprehensive investigation of the event is still needed.

Although blending affects only a minor fraction of (NEO)WISE targets, correcting it with centroid information will improve their light curves. 
This method can be used with any standard imaging sequence to get light curves for VIMs, provided their angular separation is measured by other means. A single high-resolution imaging observation using adaptive optics or interferometry might be sufficient to re-assess archival light curves or may eliminate the need for sophisticated monitoring.

\begin{acknowledgments}
This research has made use of the NASA/IPAC Infrared Science Archive, which is funded by the National Aeronautics and Space Administration and operated by the California Institute of Technology. The author thanks Neha Patak for providing the catalog before publication.
\end{acknowledgments}

\facilities{ATLAS, GAIA, IRSA, NEOWISE, {\em Spitzer}, WISE}
\software{Data analysis was done using Interactive Data Language (IDL) (Exelis Visual Information Solutions, Boulder, Colorado).}

\bibliographystyle{aasjournal-hyperref}
\bibliography{lcd}
\end{document}